\documentclass[submission, Phys]{SciPost}

\usepackage{amsmath}
\usepackage{amssymb}
\usepackage{graphicx}

\begin{document}

\begin{center}{\Large \textbf{Critical properties of a comb lattice
}}\end{center}

\begin{center}
N. Chepiga\textsuperscript{1*},
Steven R. White\textsuperscript{2}
\end{center}

\begin{center}
{\bf 1} Institute for Theoretical Physics, University of Amsterdam, Science Park 904, 1098 XH Amsterdam, The Netherlands
\\
{\bf 2} Department of Physics and Astronomy, University of California, Irvine, CA 92697, USA
\\
* natalia.chepiga@alumni.epfl.ch
\end{center}

\begin{center}
\today
\end{center}
%
%
\section*{Abstract}
{\bf In this paper we study the critical properties of the Heisenberg spin-1/2
model on a comb lattice --- a 1D backbone decorated with finite 1D chains --
the teeth. We address the problem numerically by a comb tensor network that
duplicates the geometry of a lattice. We observe a fundamental difference between
the states on a comb with even and odd number of sites per tooth, which
resembles an even-odd effect in spin-1/2 ladders. The comb with odd teeth is
always critical, not only along the teeth, but also along the backbone, which
leads to a competition between two critical regimes in orthogonal directions.
In addition, we show that in a weak-backbone limit the excitation energy scales as
$1/(NL)$, and not as $1/N$ or $1/L$ typical for 1D systems. For even teeth in the
weak backbone limit the system corresponds to a collection of decoupled
critical chains of length $L$, while in the strong backbone limit, one spin from each
tooth forms the backbone, so the effective length of a critical tooth
is one site shorter, $L-1$. Surprisingly, these two regimes are connected via a
state where a critical chain spans over two nearest neighbor teeth, with an effective
length $2L$.  }

\vspace{10pt}
\noindent\rule{\textwidth}{1pt}
\tableofcontents\thispagestyle{fancy}
\noindent\rule{\textwidth}{1pt}
\vspace{10pt}

\section{Introduction}

Antiferromagnetic  Heisenberg  spin  chains  have  been studied  intensively  over  the  years. 
In one-dimensional systems frustration is usually induced through competing
interactions and is known to lead to various exotic phases and quantum phase
transitions. The effects from competing interactions, which in 1D are often
equivalent to geometric frustration, have been studied intensively within the
framework of $J_1-J_2$
chains\cite{MajumdarGhosh,okamoto,kolezhuk_prl,kolezhuk_connectivity,PhysRevB.100.104426,2020arXiv200208982C}
and spin
ladders\cite{Dagotto,PhysRevB.84.144407,PhysRevLett.77.3443,PhysRevLett.105.077202,PhysRevB.88.184418}.
Alternatively, geometric frustration can be added to a system through a
decoration with dangling
spins\cite{igarashi,2000PhyA..286..181A,PhysRevB.81.214431}, sometimes known as
a Kondo necklace problem\cite{necklace,necklace2}. Traditionally the number of
decorating spins in the necklace models is limited to just a few. In the
present manuscript we will focus on a so-called comb lattice where the number
of pending spins is comparable to the length of the main chain - the backbone.

\begin{figure}[t!]
\centering
\includegraphics[width=0.48\textwidth]{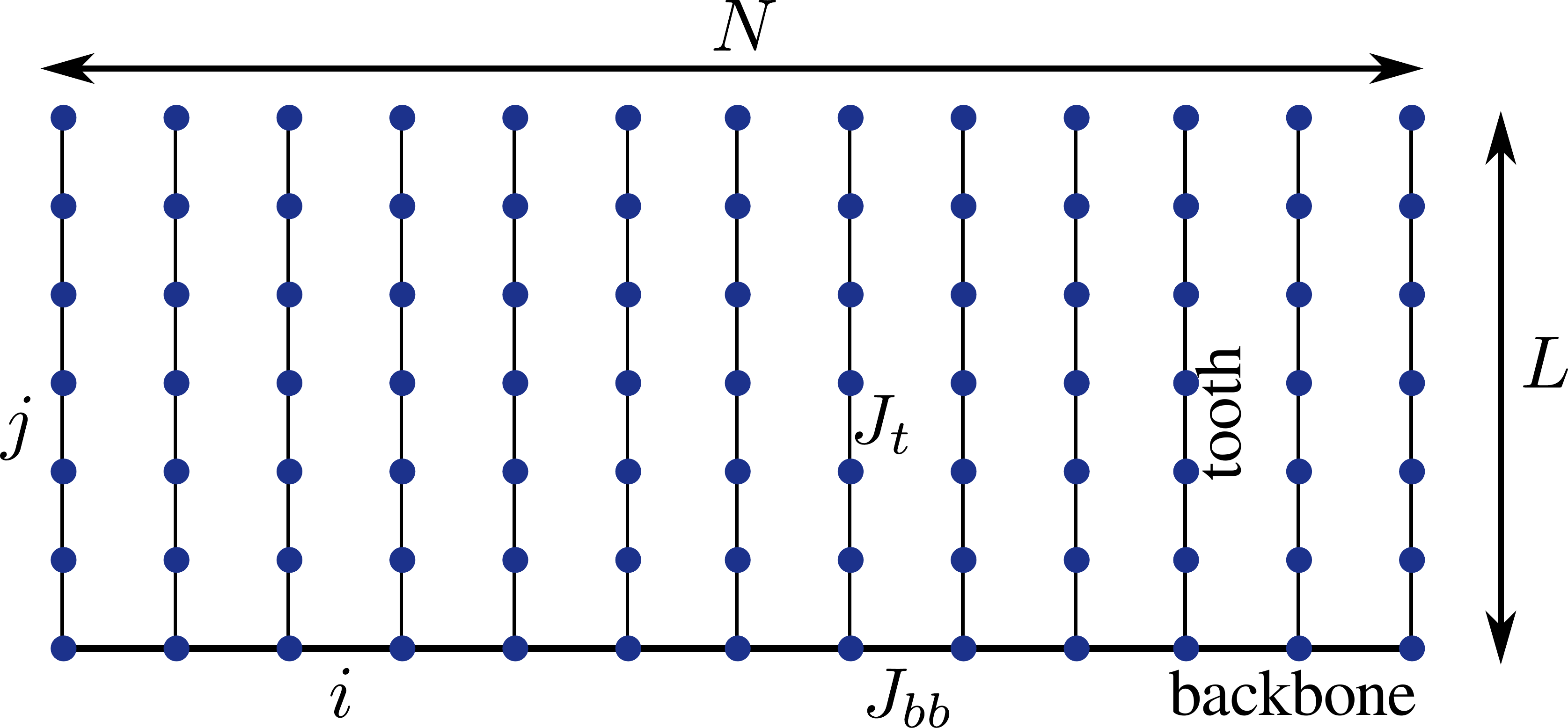}
\caption{(Color online)  Sketch of the comb lattice. Each tooth corresponds to
a finite-size spin-1/2 Heisenberg chain with nearest-neighbor coupling
constant $J_{t}=1$. Edge spins on one edge of the tooth are coupled along
the backbone with coupling constant $J_{bb}$.} \label{fig:comb_sketch}
\end{figure}

The simplest comb lattice is a tree lattice that consists of spin chains
coupled to each other through one edge, as schematically sketched in
Fig.\ref{fig:comb_sketch}. A recently proposed comb tensor network\cite{comb}
provides an efficient way to simulate this model numerically. The method has
been bench-marked on a spin-1 Heisenberg comb lattice and has revealed a number
of unusual states caused by the lattice geometry. In particular, the states include an emergent
critical spin chain formed out of spin-1/2 edge states confined at the edges of
the Haldane chains, and higher-order edge-states emergent at the two corners of
a comb lattice\cite{comb}.

Identification of the universality classes of the quantum critical lines is
of central interest in one-dimensional many-body physics. In most of the
cases the critical properties can be described by the underlying conformal
field theory (CFT).  In some rare cases, however, this is not possible due to
perturbations that drive the system out of the conformal regime. Perhaps the most
celebrated example is associated with a relevant chiral perturbation that leads
to a continuous quantum phase transition is a non-conformal Huse-Fisher
universality class\cite{HuseFisher,prl_chepiga,2020arXiv200106698C}. In the
present manuscript we will provide another example, in which the conformal
criticality is destroyed by purely geometric frustration. This geometric frustration does not involve
triangular arrangements of spins or next-neighbor couplings, which clearly cause frustration to a
N\'eel spin pattern. Instead, within a valence bond picture it stems from the fact that a spin with 
three neighbors can only form a valence bond with one of them.

In the present paper we study critical properties of the spin-1/2 Heisenberg model
on a comb lattice defined by the following Hamiltonian:

\begin{equation}
  H=J_{bb}\sum_{i=1}^{N-1}{\bf S}_{i,1}\cdot{\bf S}_{i+1,1}+
  J_t\sum_{i=1}^N\sum_{j=1}^{L-1}{\bf S}_{i,j}\cdot{\bf S}_{i,j+1},
  \label{eq:comb_hamilt}
\end{equation}
where $J_{bb},J_t>0$ are antiferromagnetic nearest-neighbor interactions along
the backbone and along the teeth, correspondingly. Without loss of generality we
set the tooth coupling constant to $J_t=1$.  By contrast to the spin-1
case\cite{comb}, the system is critical even in the absence of backbone
interaction.

Unlike an infinite-dimensional Bethe lattice\cite{Baxter:1982zz}, the comb
lattice is a one dimensional system. This conclusion can be  approached from
two different starting points. First, the comb can be viewed as a set of 1D
chains -- the teeth. When the backbone interaction is not relevant it can be
considered as a special boundary conditions (or boundary field) applied to
critical chains sitting on the teeth. In this respect, the comb lattice is a
straightforward generalization of a Y-junction of three
chains\cite{PhysRevB.74.060401} and of a chain with an impurity
bond\cite{PhysRevB.46.10866} studied previously. For the latter it has been
shown that the ground state of critical chains changes drastically while tuning
the impurity coupling. On the other hand, the comb is a highly decorated chain -- the backbone; 
and can be considered as a generalization of the necklace
problem.  In this paper we study what happens in the interplay of these two
regimes when the system is critical, in particular, when it is critical in {\it
in both directions}.

\begin{figure}[t!]
\centering
\includegraphics[width=0.85\textwidth]{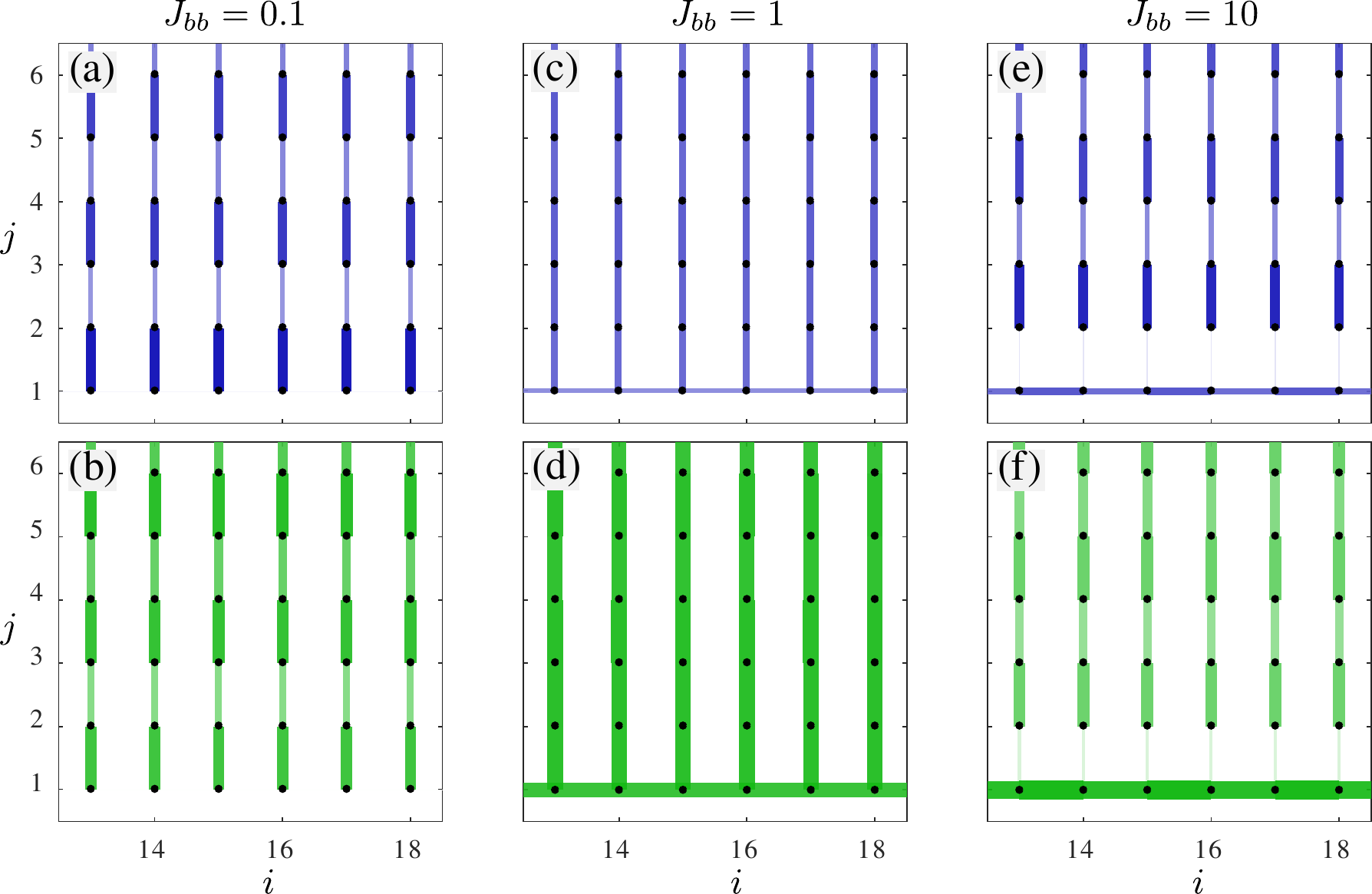}
\caption{(Color online) Nearest-neighbor spin-spin correlations (upper panels)
and the entanglement entropy (lower panels)  on a comb lattice with various
coupling constants on a  backbone. Only a small part in the middle of the
backbone of a comb with $30\times 30$ sites is shown. The width of the
lines and the intensity of the color are proportional to the strength of
the correlation (upper panels, blue) or entanglement (lower panels, green).
}
\label{fig:graph}
\end{figure}

\begin{figure}[t!]
\centering
\includegraphics[width=0.85\textwidth]{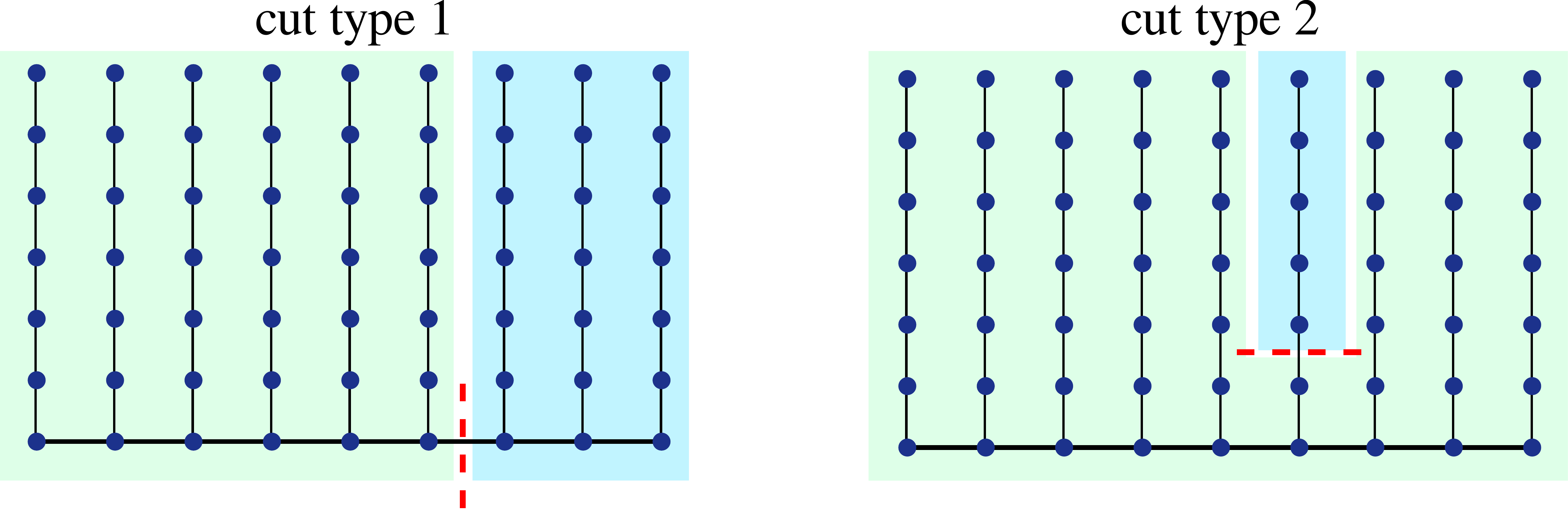}
\caption{(Color online) Two types of bipartition used in this work to compute the entanglement entropy, background colors mark two subsystems created upon the bipartition. 
}
\label{fig:bipartition}
\end{figure}

In our study we will mainly focus on the three regimes: a weak backbone limit
$J_{bb}\ll 1$, a competition between the teeth and the backbone $J_{bb}\approx
1$, and a strong backbone regime $J_{bb}\gg 1$. Fig.\ref{fig:graph} provides a
first insight into these three cases. It shows nearest-neighbor correlations
(blue) and bipartite entanglement entropy (green) on nearest-neighbor bonds.
In order to compute the entanglement entropy we divide the system into two different pieces in two different ways, as shown in Fig.\ref{fig:bipartition}, and compute the reduced density matrix $\rho$. The entanglement entropy is then given by $S=-\mathrm{Tr} \rho \ln \rho$.
 In the first type of bi-partition, the system is cut across the backbone such that each tooth belongs entirely to one of the two subsystems. In this way we measure the entanglement carried by the backbone. In the second type of bi-partition, one subsystem includes a set of sites at the tip of the selected tooth, while another subsystem contains all the remaining part of the comb, as shown in the right panel of Fig.\ref{fig:bipartition}. 
 
In 
order to focus on bulk behavior, we look at a small window in the
middle of the backbone of a $30\times 30$ comb. When $J_{bb}=0.1$
the correlations and the entanglement are concentrated within the teeth, which
remain almost uncorrelated and disentangled. One sees also a dimerization
that appears at the end of each tooth and associated Friedel oscillations. This is a
common consequence of open boundary conditions in critical Heisenberg chains.
At intermediate backbone couplings, specifically $J_{bb}\approx 1$, the
nearest-neighbor correlations along the tooth and along the backbone are almost
equal. Moreover, the entanglement is almost equally distributed on all the
bonds not too far from the backbone. In practice, this means that cutting, say
3/4 of one tooth, or a half of the entire comb cost essentially the same amount
of entanglement. In the third regime the backbone coupling is so strong that
the system prefers to have as much correlation and entanglement within the
backbone. As a results, the strong backbone is almost completely decoupled from
the rest of the system. This naturally makes weakly coupled teeth one site
shorter. In the following we will provide more details on each of these three
regimes.

\section{The weak-backbone limit}

Let us first consider the limit of a weak backbone interaction. According to
Fig.\ref{fig:graph} we might expect nearly decoupled Heisenberg chains on
teeth. An isolated spin-1/2 Heisenberg chain is known to be described by the
Wess-Zumino-Witten (WZW) SU(2)$_{k=1}$\cite{affleck_haldane} critical conformal
field theory (CFT) in 1+1D. According to WZW SU(2)$_1$ CFT the singlet-triplet
gap scales with the length of a chain $L$ as $E_1-E_0=\pi v/L$, up to
some logarithmic corrections. In the limit of nearly decoupled teeth the
singlet-triplet excitation on the comb lattice can also be considered as a
corresponding excitation of a single tooth. In the presence of a weak
inter-tooth interaction one naturally expects this excitation to be slightly
delocalized, corresponding to a finite-width soliton along the backbone.
Interestingly enough, the singlet-quintuplet excitation of an isolated chain
scales with its length as  $E_2-E_0=4\pi v/L$. It makes it
energetically favorable for a comb to accommodate several spin-1 solitons
before exciting a tooth to a higher state.
 
\begin{figure}[t!]
\centering
\includegraphics[width=0.95\textwidth]{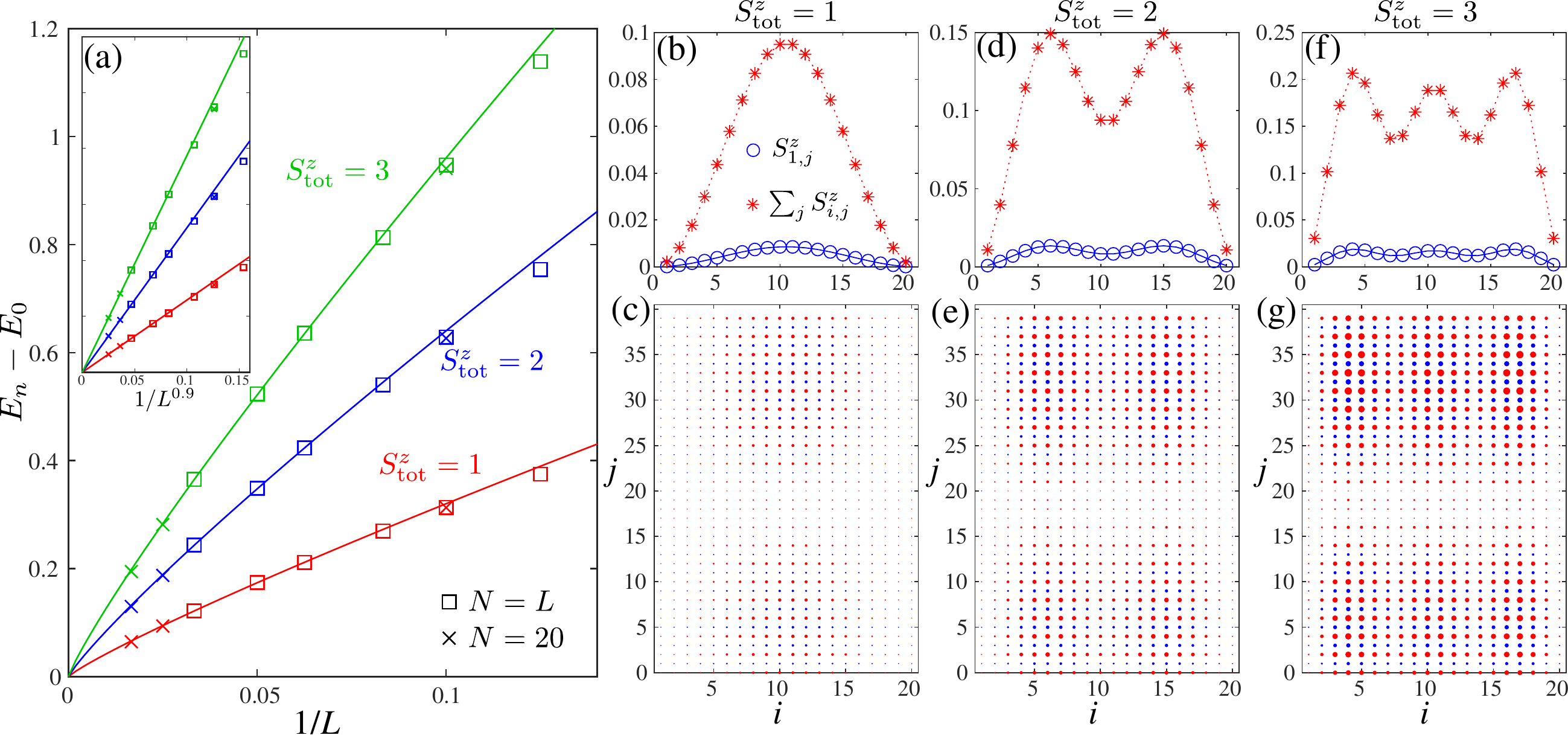}
\caption{(Color online) Magnetic excitations in a weak-backbone regime. (a)
Scaling of the energy difference between the lowest energy state in the
sector of total magnetization $S^z_\mathrm{tot}=1,2,3$ and the ground-state
in the sector $S^z_\mathrm{tot}=0$ as a function of the inverse of the tooth
length $1/L$. The symbols are the DMRG data, the red line is the result of the fit
of the singlet-triplet gap with $\propto L^{-d_\mathrm{app}}$, and the blue (green) line is the
result of the fit multiplied by an integer 2 (3). Inset: The same plot, but
as a function of $1/L^{0.9}$. (b-g) The distribution
of local magnetization on a comb lattice with $L=40$ and $N=20$. (b), (d),
(f) The local magnetization on the backbone (blue circles) and total
magnetization of a tooth as a function of tooth index (red stars). (c),
(e), (g) Local magnetization, with the size of the circles proportional to the
absolute value of the magnetization, and with red and blue colors signifying positive and
negative signs of magnetization.} \label{fig:gap_01_even}
\end{figure}

This picture is confirmed by our numerical results presented in
Fig.\ref{fig:gap_01_even}(a). By looking at the excitation energy between the
lowest state in the sectors with different total magnetization we observe 
states with one, two and three magnetic solitons. It is spectacular that the
finite-size scaling of the two- and three- solitons state are in perfect
agreement with the scaling for a single soliton multiplied by a corresponding
integer. In practice, this means that on a chosen scale, these solitons remain
almost decoupled. 

The scaling deviates from linear behavior due to the presence of logarithmic corrections that for an isolated chain take the form $\propto-\frac{\pi v}{L\log L}$. Therefore, an apparent scaling dimension $d_\mathrm{app}$ extracted from the fit to $E_n-E_0=\pi v n /L^{d_\mathrm{app}}$ is expected to be smaller than its true value $d=1$. This qualitatively agrees with our finding shown in Fig.\ref{fig:gap_01_even}(a) that $d_\mathrm{app}\approx 0.9$. Another source of log-corrections is caused by a weakening of the boundary conditions at one edge of a tooth due to a presence of backbone interactions; although according to Fig.\ref{fig:friedel}(a) this boundary effect is relatively small for $J_{bb}\lesssim 0.3$.
 
Finally, we would like to stress the
independence of the results on the number of teeth in the comb. Of course, this
statement is true only when the number of solitons is sufficiently small
compare to the total number of teeth; however even for three solitons allocated
on a comb with $N=8$ teeth this property holds reasonably well.

So far we have only considered a comb with an even number of sites per tooth,
but the picture changes drastically when $L$ is odd. An isolated Heisenberg
chain with an odd number of sites has total spin $S=1/2$. Therefore, as soon as
the backbone interaction is non-zero the spin-1/2 degrees of freedom on each
tooth form a critical spin-1/2 chain. This can be detected, in particular, by
looking at the Friedel oscillation profile along the backbone. 

\begin{figure}[t!]
\centering
\includegraphics[width=0.6\textwidth]{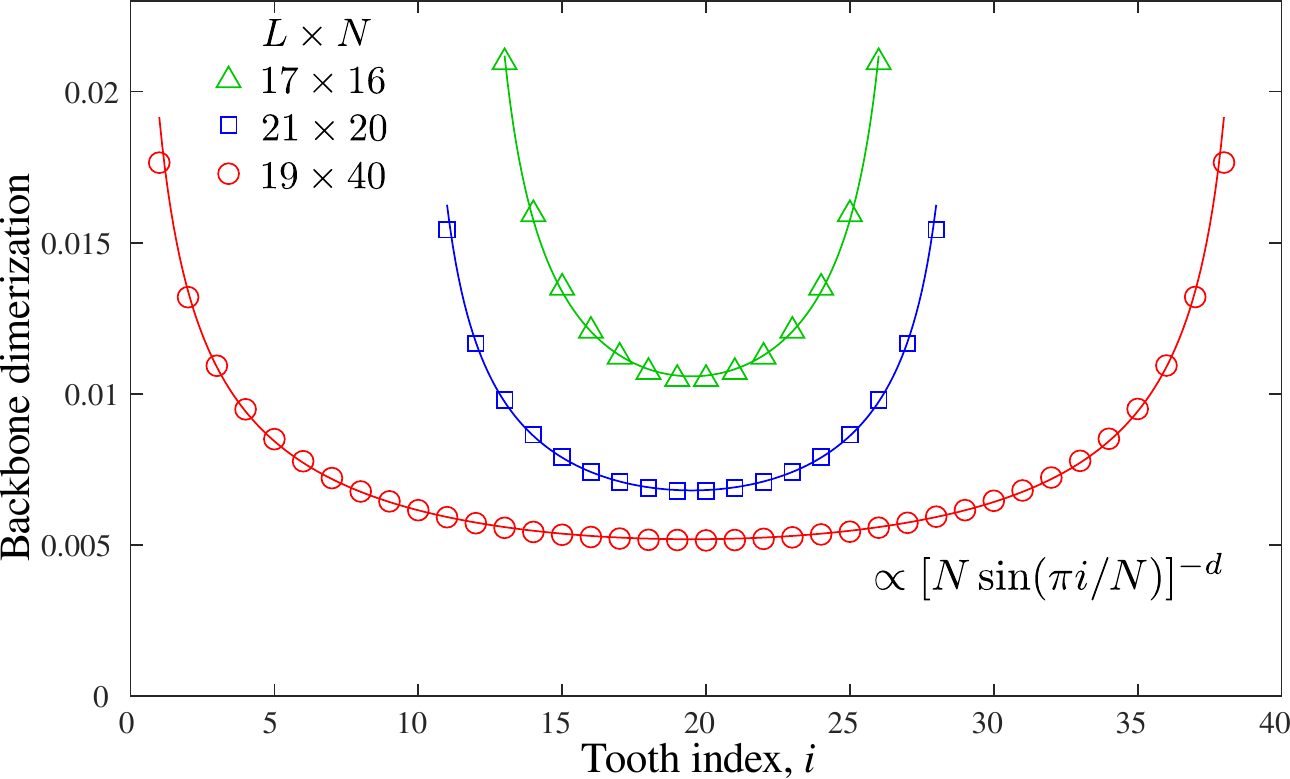}
\caption{(Color online) Friedel oscillations along the backbone in a comb with
an odd number of sites per tooth $L$. Symbols are our numerical data, lines
are fit to the CFT prediction for the envelope. The resulting critical exponents
are $d\approx0.44$ for $17\times16$, $d\approx0.48$ for $21\times20$, and
$d\approx0.51$ for $19\times40$, which are in good agreement with the CFT
prediction $d=1/2$.  } \label{fig:fried_01_odd}
\end{figure}

In an isolated spin-1/2 chain the open edges favor dimerization. This acts as fixed boundary conditions and induces Friedel
oscillations. According to the boundary conformal field theory\cite{1991PhLB..259..274C} at the critical point the dimerization scales away from the boundary as $D\propto x^d$, where $x$ is the distance to the boundary and $d$ is the corresponding scaling dimension. For a finite chain with $N$ sites the conformal transformation gives:
\begin{equation}
  D(i)\propto \frac{1}{[(N/\pi)\sin(\pi i/N)]^d},
\end{equation}
where $D(i)=|\langle {\bf S}_{i-1,1}\cdot{\bf S}_{i,1}\rangle-\langle{\bf S}_{i,1}\cdot{\bf S}_{i+1,1}\rangle|$ 
is the dimerization on the backbone. 
In the WZW SU(2)$_k$ critical theory the dimerization is induced by a $j=1/2$ operator with scaling dimension $d=2j(j+1)/(k+2)$\cite{AffleckGepner}. So, for the critical spin-1/2 chain described by WZW SU(2)$_1$ an expected scaling dimension is $d=1/2$.
Fits of our data for different clusters are shown in Fig.\ref{fig:fried_01_odd}. 
The numerically extracted values of the critical exponent $d\approx 0.51$ for
$19\times 40$ and $d\approx 0.48$ for $21\times 20$ are in a good agreement
with the CFT prediction $d=1/2$. It is important that even at the edges the
dimerization is very small. This is because every spin-1/2 sitting on the backbone is
mainly involved in the formation of a stronger dimer along the tooth each of
which also induces Friedel oscillations perpendicular to the backbone.

\begin{figure}[t!]
\centering
\includegraphics[width=0.95\textwidth]{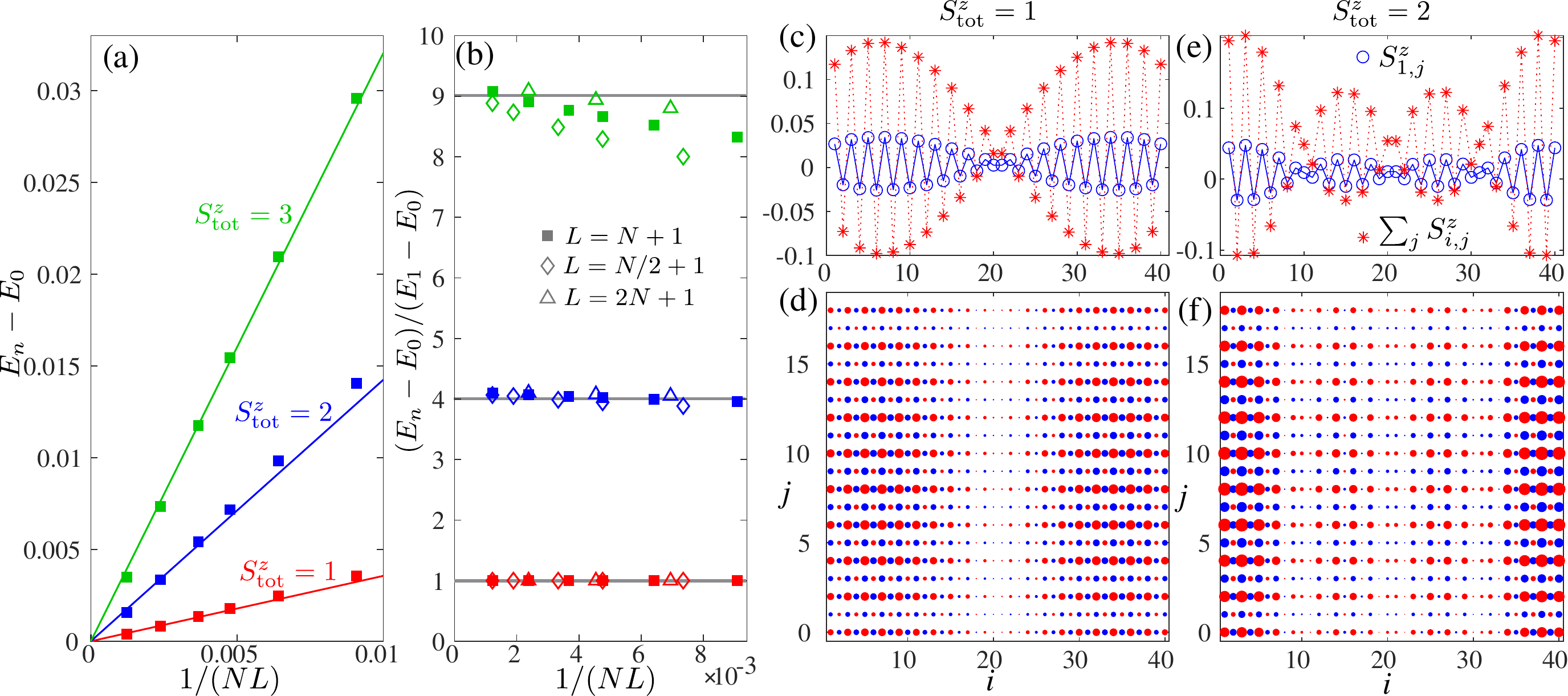}
\caption{(Color online) Magnetic excitations on a comb lattice with an odd number
of sites per tooth in the limit $J_{bb}\ll 1$. (a) Finite size scaling of
the energy difference between the singlet ground-state and the lowest
triplet (red), quintuplet (blue) and septuplet (green) states.
Lines are linear fit of the singlet-triplet gap as a function of $1/(NL)$  multiplied by the
expected structure of conformal tower $n=1,4,9$. (b) The conformal tower of
states extracted as a ratio of the gaps for each fixed size of the comb.
Results for $n=1$ are trivial and shown for completeness. (c)-(f)
Distribution of local magnetization on a comb lattice with $L=19$ and
$N=40$.  (c), (e) Local magnetization on the backbone (blue circles) and
total magnetization of a teeth as a function of tooth index $i$ (red
stars). (d), (f) Local magnetization, with the size of the circles proportional
to the absolute value of magnetization; red and blue colors indicate 
positive and negative signs of magnetization} \label{fig:gap_01_odd}
\end{figure}

\begin{figure}[t!]
\centering
\includegraphics[width=0.75\textwidth]{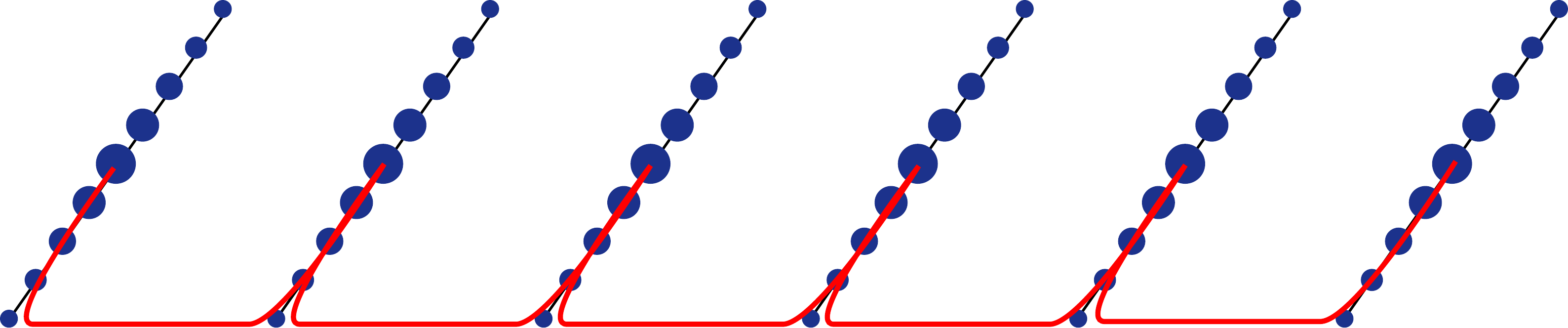}
\caption{(Color online) Schematic representation of a state in a comb with an odd
number of sites per tooth. Each tooth is in the state with $s=1/2$, which
is delocalized along the tooth. When the backbone interaction is
sufficiently small, a maximal probability to find spin-1/2 is approximately
in the middle of {\it each} tooth. So the distance between two spin-1/2
degrees of freedom is proportional to $L$ and the entire length of the
effective chain is proportional to $(NL)$. Upon increasing the backbone
interaction, this maximum of the spin-1/2 profile on a tooth moves towards
the backbone but in a non-uniform way.  }
\label{fig:delocalized}
\end{figure}

Let us now look at the excitation spectrum of a comb with odd tooth length in the weak backbone regime. 
Quite surprisingly, the energy gap scales linearly with $1/(NL)$, as shown in Fig.\ref{fig:gap_01_odd}(a).
This can be justified by a simple argument illustrated in
Fig.\ref{fig:delocalized}. Each tooth of a comb acts as an effective spin-1/2
degrees of freedom delocalized along the tooth. When the backbone interaction
is not too big the maximum of magnetization profile on {\it each} tooth is
approximately in its middle. So the effective distance between the nearest
spin-1/2 degrees of freedom is proportional to $\propto L$, and more
importantly does not change much along the chain. Then the total length of an
effective spin chain is proportional to $(NL)$. According to the CFT, the
energy gap scales linearly with the effective length of a critical chain, which
implies $\propto (NL)^{-1}$.

According to boundary CFT an excitation energy of a chain with conformally-invariant boundary conditions scales with the length of the chain $\tilde{L}$ as $\pi n v/\tilde{L}$, where $v$ is a non-universal sound velocity and $n$ is a numerical factor associated with the energy levels  that belongs to the so-called conformal towers of states.
The structure of the low-energy spectra for WZW SU(2)$_k$ models with specified total spin has been worked out by Affleck et al.\cite{AffleckGepner} by means of conformal field theory.
Numerical calculation of {\it all} low-lying energy levels is often a challenging and computationally demanding task. By contrast, it has long been known that energy of magnetic excitations including singlet-triplet and singlet-quintuplet gaps can easily be obtained with DMRG by converging the lowest energy states within different sectors of U(1) symmetry, or in other words, with different total magnetization $S^z_\mathrm{tot}=0,1,2,$...
 This method does not provide the full low-energy spectra, including multiplicities of the energy levels, but only the outer envelope. However, the special structure of the envelope is often sufficient to distinguish between various CFT candidates\cite{PhysRevB.93.241108,2020arXiv200208982C}. It turns out, that for the WZW SU(2)$_1$ critical point with zero-spin ground-state, realized in spin-1/2 chains with even number of sites,  the form of the envelope is exceptionally simple: the singlet-triplet gap scales as $(\pi v)/\tilde{L}$, the singlet-quintuplet as $(4\pi v)/\tilde{L}$, and the singlet-septuplet as $(9\pi v)/\tilde{L}$. 

The results
presented in Fig.\ref{fig:gap_01_odd}(a) for a square comb with $N=L-1$,
exhibit a good agreement with linear scaling with $1/(NL)$. This implies that in an effective CFT the linear measure of the system is given by $\tilde{L}\propto NL$.
 In order to check the
$n=1,4,9$-structure we fit the lowest singlet-triplet gap to $\pi v/(NL)$, and
multiply the obtained linear function with $n=4$ (blue line) and $n=9$ (green)
which both are in excellent agreement with the data points. In
Fig.\ref{fig:gap_01_odd}(b) we extract the value of the slope pre-factor $n$
numerically and compare it with the CFT $1,4,9$ prediction. Note that it has
been obtained by dividing various magnetic gaps by the singlet-triplet gap, so
results for $n=1$ are trivial. In addition, by looking at the distribution of the
local magnetization shown in Fig.\ref{fig:gap_01_odd}(c)-(f) one can
immediately recognize the butterfly profile distinct for the critical chains
and so different from the results obtained for the comb with even number of
sites per tooth.

To summarize, the ground state properties of a spin-1/2 comb with even and odd
teeth are fundamentally different.  In a weak backbone limit $J_{bb}\ll 1$, the
comb with even length teeth tends to screen the effect of the backbone interaction,
while the comb with odd length teeth, as soon as the backbone coupling is non-zero,
corresponds to the critical spin-1/2 chain in the direction of a backbone. Here
one can intuitively rely on an analogy with spin ladders: when the number of
legs is even each rung is in the $j=0$ state; when the number of legs is odd the rungs are in $j=1/2$
states. This gives the celebrated conclusion that spin-1/2 ladders with even
number of legs are gapped, and it is gappless if the number of legs is odd.
Moreover, the delocalized nature of the spin-1/2 degrees of freedom in combs with
odd teeth lead to a very unusual (for one dimension) scaling of the excitation
spectra - linear scaling with $1/(NL)$. The low-energy physics of combs with
mixed even and odd teeth can be guessed based on these conclusions but the
detailed numerical investigation of such mixed systems is beyond the scope of
this paper.

\section{Formation of double-teeth chains}

Let us now look what happens to the ground state when the backbone interaction
is tuned from $J_{bb}<<1$ to $J_{bb}\approx 1$ in a comb with even teeth. As
shown above in the regime with small backbone coupling, the teeth are rather
independent from each other and correspond to the critical Heisenberg chain
described by WZW SU(2)$_1$ conformal field theory. In case of isolated
chains, open boundary conditions favor dimerization and act as a fixed boundary
condition, which in turn induce strong Friedel oscillations. We have
already discussed the profile of the Friedel oscillation along the backbone in the context of a
comb with odd teeth and weak backbone interaction. The profile of the Friedel oscillations along the isolated tooth is given by a similar expression:

\begin{equation}
  D(j)\propto \frac{1}{[L\sin(\pi j/L)]^d},
\end{equation}
where $D(j)=(-1)^j\left[\langle{\bf S}_{i,j-1}\cdot{\bf S}_{i,j} \rangle-\langle{\bf S}_{i,j}\cdot{\bf S}_{i,j+1} \rangle\right]$ 
is the dimerization, $1\leq i\leq N$ is a tooth index, 
$1\leq j \leq L$ is a site index within the tooth. 

\begin{figure}[t!]
\centering
\includegraphics[width=0.7\textwidth]{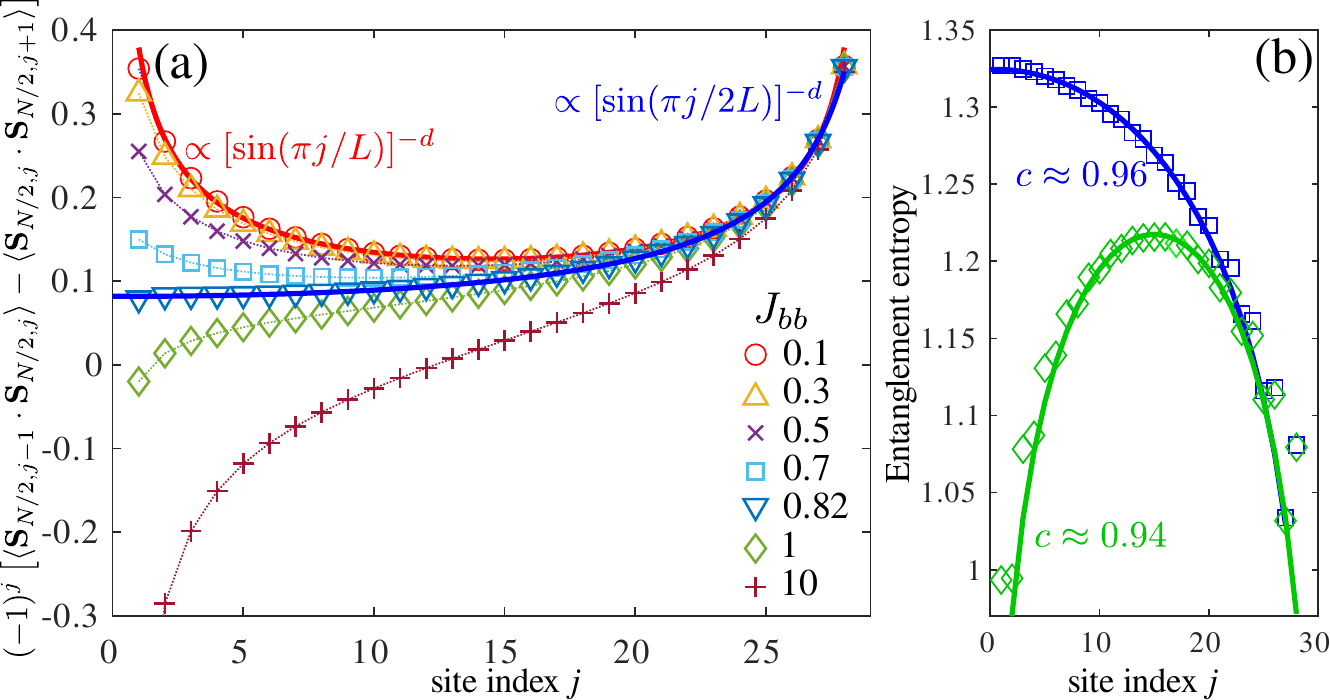}
\caption{(Color online) (a) Friedel oscillations along the middle tooth for
various backbone couplings in a comb with $L=30$ sites per tooth  and
$N=30$ teeth. Red and blue lines corresponds to the CFT fit with chain
length $L$ and $2L$ correspondingly. (b) Entanglement entropy profile along
the middle tooth for $J_{bb}=0.1$ (green) and $0.82$ (blue); lines are fit
to the Calabrese-Cardy formula.  }
\label{fig:friedel}
\end{figure}

Fig.\ref{fig:friedel}(a) presents the Friedel oscillations along the middle
tooth $i=N/2$ for various values of the backbone interaction for a comb with
$L=30$ and $N=30$. For small values of $J_{bb}=0.1$ the shape of the envelope
is well captured (red symbols and line) and the scaling dimensions obtained
numerically $d\approx 0.49$ is in excellent agreement with the CFT prediction.
Slight increase of the backbone interaction smears down fixed boundary
conditions and therefore suppress the Friedel oscillations at the edge
connected to the backbone. The effect is very small for $J_{bb}=0.3$, but
starting from  $J_{bb}\approx 0.5$ the deviation from the CFT profile is
significant.

We noticed that further increasing the backbone interaction drives the comb
though a point where the oscillation profile on a tooth resembles  half of the
CFT envelop on a chain with double length. In Fig.\ref{fig:friedel}(a) this
happens at $J_{bb}\approx0.82$. By fitting the Friedel oscillations to the
(twice as big) envelop $D(j)\propto \frac{1}{[\sin(\pi (j+L)/(2L))]^d}$ we find
very good agreement with our data points, although the critical exponent
extracted from this fit $d\approx 0.65$ is quite far from the theoretical value
$d=1/2$. 

\begin{figure}[t!]
\centering
\includegraphics[width=0.4\textwidth]{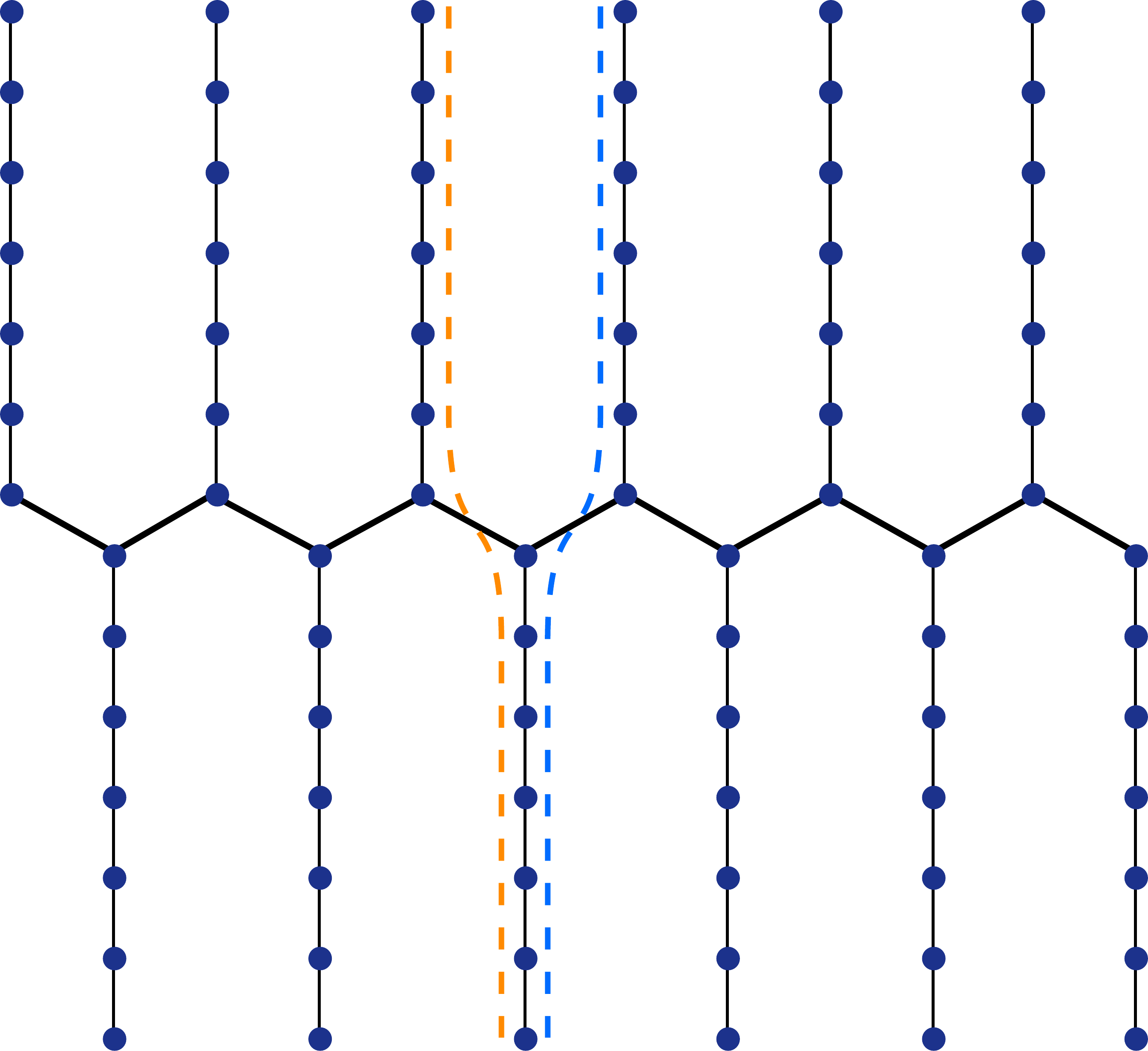}
\caption{(Color online) Alternative representation of a comb geometry. When the
backbone interaction is comparable to the interaction within the teeth, the
system can be viewed as a collection of one-dimensional chains (dashed
lines) with the distortion in the middle of the chain.  }
\label{fig:LongChains}
\end{figure}

This observation suggests that when the coupling along the backbone and along
the tooth are comparable $J_{bb}\approx J_t$ the comb resembles the collection
of 1D chains extended over two neighboring teeth as sketched in
Fig.\ref{fig:LongChains}.

We can extract the central charge from the entanglement entropy profile along a middle tooth $i=N/2$.
According to Calabrese-Cardy formula, in a finite-size chain with $L$ sites and open boundary
conditions, the entanglement entropy scales with the size of the subsystem $l$ as\cite{CalabreseCardy}:
 
\begin{equation}
\tilde{S}_L(l)={S}_L(l)-\zeta\langle {\bf S}_{i,j}{\bf S}_{i,j+1}\rangle=\frac{c}{6}\ln \frac{2L}{\pi}\sin\frac{\pi l}{L}+s_1+\log g,
\end{equation}
where $c$ is a central charge and $\zeta\approx 1$ is a non-universal constant
used to suppress the Friedel oscillations\cite{capponi} when the system is cut across the bond $\{(i,j),(i,j+1)\}$ and $s_1$ and $\log g$ are non-universal and universal constants.   First, we benchmark the results for
$J_{bb}=0.1$. The fit of our numerical data along the central tooth to the
Calabreze-Cardy formula gives a central charge $c\approx0.94$, which is in a
decent agreement with the CFT prediction $c=1$. 
At $J_{bb}=0.82$ when with the Friedel oscillations we observe the resemblance of 
a double-teeth chain, we find that the entanglement entropy profile also looks
like a half of the profile expected for a chain with $2L$ sites. The result
of the fit is in excellent qualitative and quantitative agreement with $c=1$
profile on a chain of length $2L$.

In the simplest case of a comb with two teeth the formation of a double-teeth
chain is exact and has been studied by Affleck and
Eggert\cite{PhysRevB.46.10866}. One can identify the following three regimes:
When the coupling constant at the  impurity is lower than the one in the bulk, the
system renormalizes to two decoupled chains each of size $L$. When the coupling
at an impurity is stronger than the coupling in the bulk, two spins connected
by an impurity bond form a singlet  and are effectively decoupled from the
remaining chains of size $L-1$ each. Finally, when there is no impurity, i.e.
when the coupling on a selected bond is equal to the coupling in the bulk, the
system is  equivalent to a  chain with $2L$ sites.  In a comb with multiple
teeth, restoration of the $2L$ chain is not expected to be exact, since each spin
located at the backbone has a coordination number three and not two. That is
why the agreement with the CFT profiles for chains with $2L$ sites shown in
Fig.\ref{fig:friedel} is impressive.
As a final remark, let us point out that upon further increase of the backbone
interaction we observe the third regime discussed by Affleck and Eggert: the
backbone chain is decoupled from the teeth each of which is one site shorter
than before. In particular, it implies that the Friedel oscillation profile for
$J_{bb}=10$ shown in Fig.\ref{fig:LongChains}(a) is antisymmetric as in the
chains with odd number of sites.

\section{Excitations at $J_{bb}=1$}

\begin{figure}[t!]
\centering
\includegraphics[width=1\textwidth]{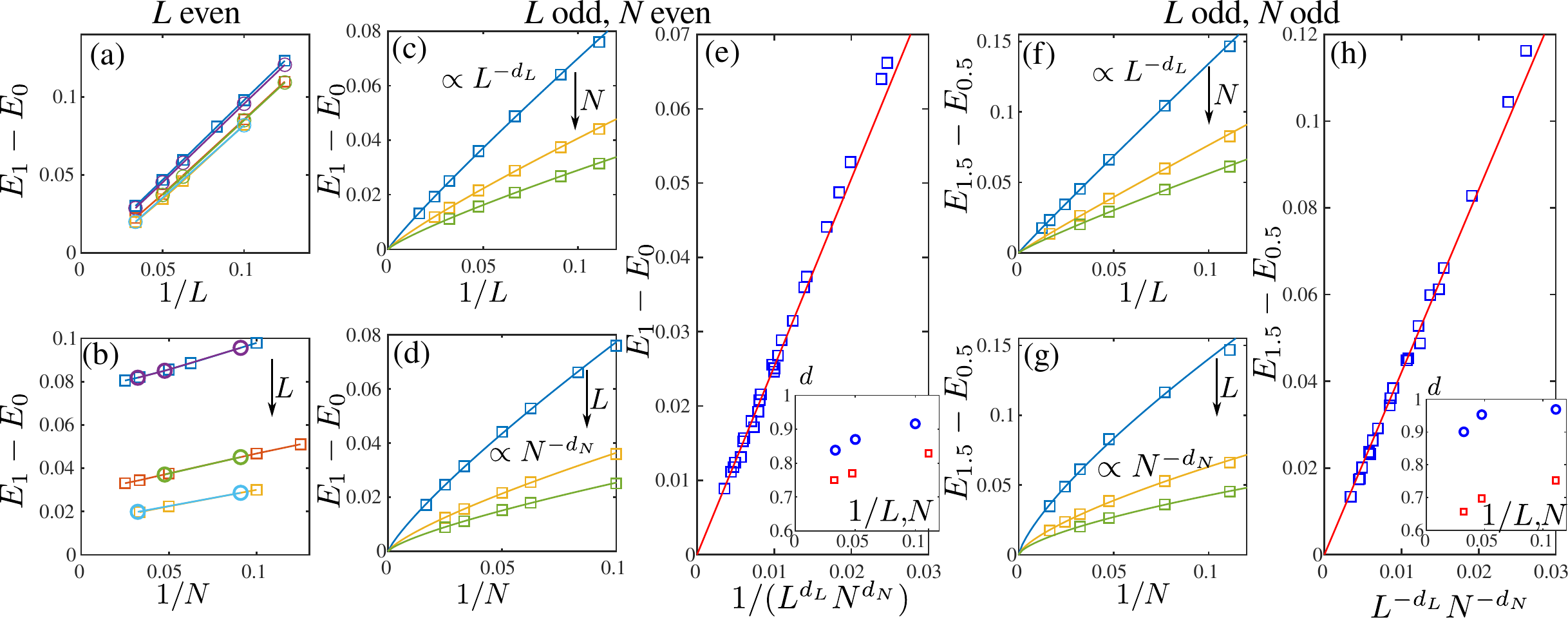}
\caption{(Color online) Finite-size scaling of energy gaps to the lowest
magnetic excitation at $J_{bb}=1$. Finite size scaling is shown separately
as a function of $1/L$ in (a),(c),(f) and as a function of $1/N$ in
(b),(d),(g). In (a)-(b) we show results for $L$-even and $N$ either
even (circles) or odd (diamonds). Panels (c)-(e) show results for
$L$-odd and $N$-even, while panels (f)-(h) are for both $N,L$-odd.  The lines in
panels (c),(f) are results of a fit of the form $\propto L^{-d_L}$, and
in panels (d),(g) of the form $\propto N^{-d_N}$. Panels (e) and (h) show
the data collapse for the best available pair of ($d_N,d_L$). The values of
$d_{N,L}$ are summarized in the insets of panels (e) and (h).  The different
colors on panels (a)-(d),(f)-(g) correspond to different widths/lengths of a
comb.  }
\label{fig:gap_1}
\end{figure}

Now lets us take a closer look at the nature of the excited states when the backbone
and tooth coupling are comparable. For simplicity we take $J_{bb}=J_t=1$.  In
order to extract the energy gap we compute the energy of the ground-state and
the lowest energy state in the sector of $S^z_\mathrm{tot}=1$
($S^z_\mathrm{tot}=3/2$ for $N,L$ odd). The results obtained for various values
of $N$ and $L$ are summarized in Fig.\ref{fig:gap_1}.

\begin{figure}[t!]
\centering
\includegraphics[width=1\textwidth]{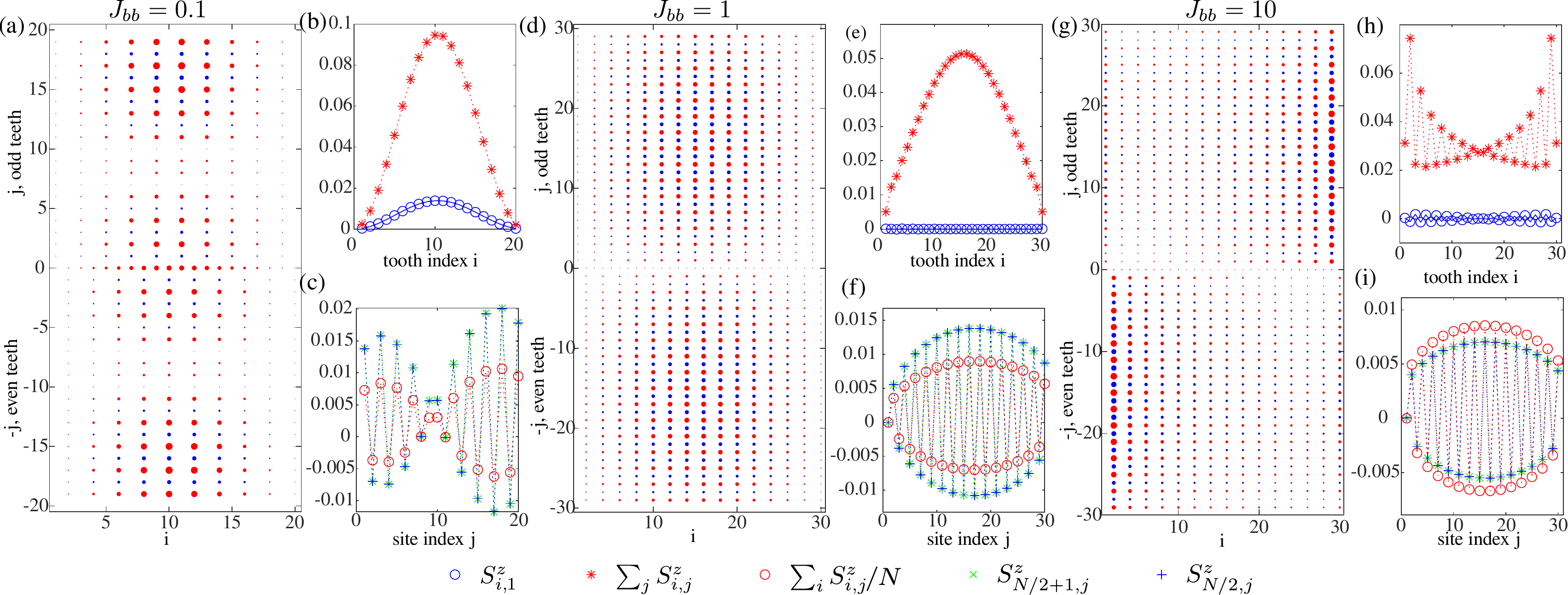}
\caption{(Color online) Local magnetization profile on combs with even $L,N$
for three different values of backbone interaction: (a)-(c) $J_{bb}=0.1$
and $L=N=20$; (d)-(f) $J_{bb}=1$ and $L=N=30$; (g)-(i) $J_{bb}=10$ and
$L=N=30$ . In (a),(d),(g) the comb is unfolded as in
Fig.\ref{fig:LongChains}; backbone sites are placed at the  coordinates
$(i,j=0)$; the radius of circles is proportional to the absolute value of
local magnetization; red and blue colors indicate positive and negative
sign of magnetization. Panels (b),(e),(h) show local magnetization on the
backbone (blue circles) and total magnetization of teeth (red stars).
Panels (c),(f),(i) show local magnetization along two consecutive teeth in
the middle of a comb (green and blue crosses), and average magnetization
profile along the teeth (red circles) } \label{fig:locmag_30_30}
\end{figure}

\begin{figure}[t!]
\centering
\includegraphics[width=1\textwidth]{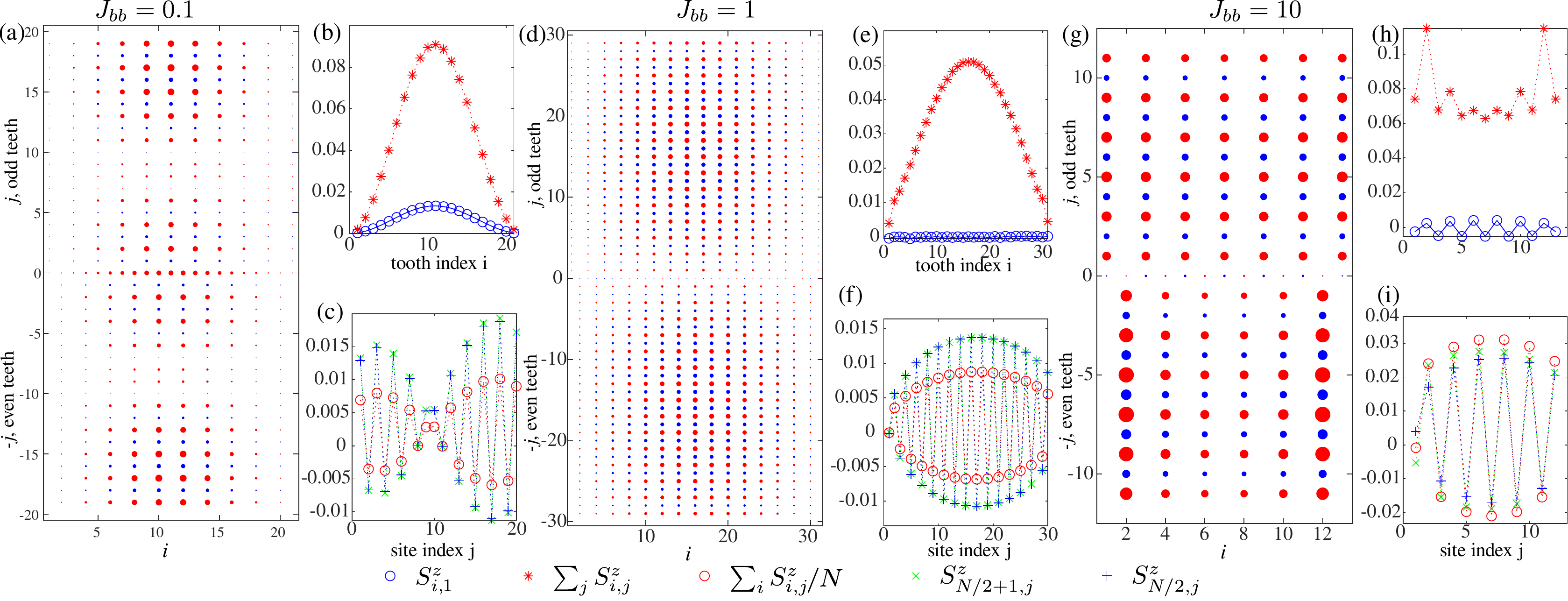}
\caption{(Color online) Same as Fig.\ref{fig:locmag_30_30} but for $L$-even and
$N$-odd: (a)-(c) and $L=20$, $N=21$; (d)-(f) $L=30$, $N=31$; (g)-(i)
$L=12$, $N=13$} \label{fig:locmag_30_31}
\end{figure}

For a comb with even number of sites per tooth  $L$ the energy of magnetic
excitations scales to zero almost linearly with $1/L$ and shows fairly small
finite-$N$ dependence. Note, that the even-odd-$N$ effect is negligible in this
case. In order to understand the nature of these excitations we plot local
magnetization profiles for both $N$-even (Fig.\ref{fig:locmag_30_30}) and
$N$-odd (Fig.\ref{fig:locmag_30_31}). For completeness we also include the
profiles in the weak- and strong-backbone limits.

\begin{figure}[t!]
\includegraphics[width=1\textwidth]{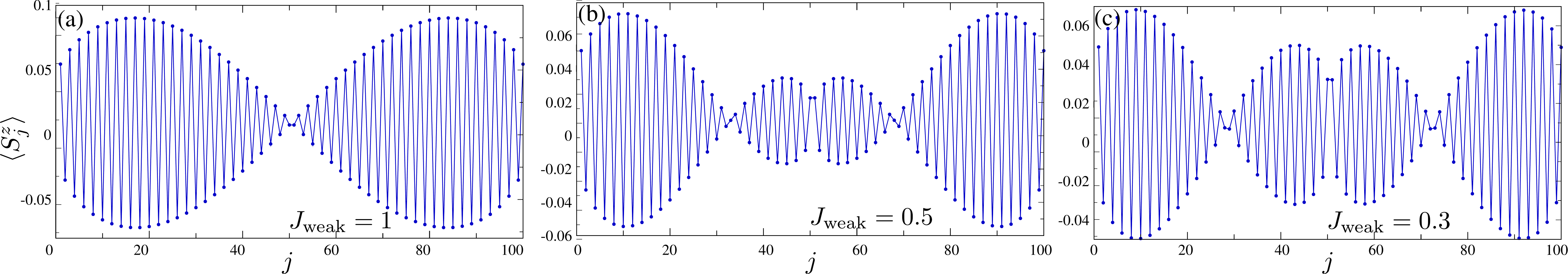}
\caption{(Color online) Local magnetization along a finite-size chain at
various values of the coupling constant at the middle bond. (a) When the
coupling is uniform along the chain, the magnetization profile has a
butterfly shape. When the coupling on the middle bond is smaller, the
double-butterfly structure emerges. The shape of the magnetization profile
at each half chain - non symmetric butterfly - is similar to the profile
observed in a comb.  }
\label{fig:weak_chain}
\end{figure}

For $L$-even and in the weak backbone limit we observe a butterfly profile
characteristic of isolated critical spin chains with even number of sites. 
Notice on Fig.\ref{fig:locmag_30_30}(c) and Fig.\ref{fig:locmag_30_31}(c)
that the butterfly is not symmetric. This can be understood by looking at the
magnetization profile on a chain with  $2L$ sites and a weak-bond impurity. The
profiles for impurity coupling have been obtained with the standard
DMRG\cite{dmrg1,dmrg2} and are summarized in Fig.\ref{fig:weak_chain}. Indeed
the profile shown in Fig.\ref{fig:weak_chain}(c) looks similar to the profiles
observed along the teeth in a weak-backbone regime in
Fig.\ref{fig:locmag_30_30}(c) and Fig.\ref{fig:locmag_30_31}(c). We also 
stress that these excitations are well localized in the backbone
direction as shown in Fig.\ref{fig:locmag_30_30}(b) and
Fig.\ref{fig:locmag_30_31}(b), which agrees with our picture of solitons. 

The profile changes significantly at $J_{bb}=1$. Along a tooth the butterfly
structure is replaced by its half, which supports the formation of a critical
chains over two-consecutive teeth discussed above. The finite-$N$ scaling shown in
Fig.\ref{fig:gap_1}(b) suggests that for any fixed and finite value of $L$ the
gap, and therefore the correlation length along the backbone remains finite.
This implies that the profile observed in Fig.\ref{fig:locmag_30_30}(b) and
Fig.\ref{fig:locmag_30_31}(b) will look like a soliton on a large-$N$ scale. In
the present case, however this soliton is perturbed by boundaries.

\begin{figure}[t!]
\centering
\includegraphics[width=1\textwidth]{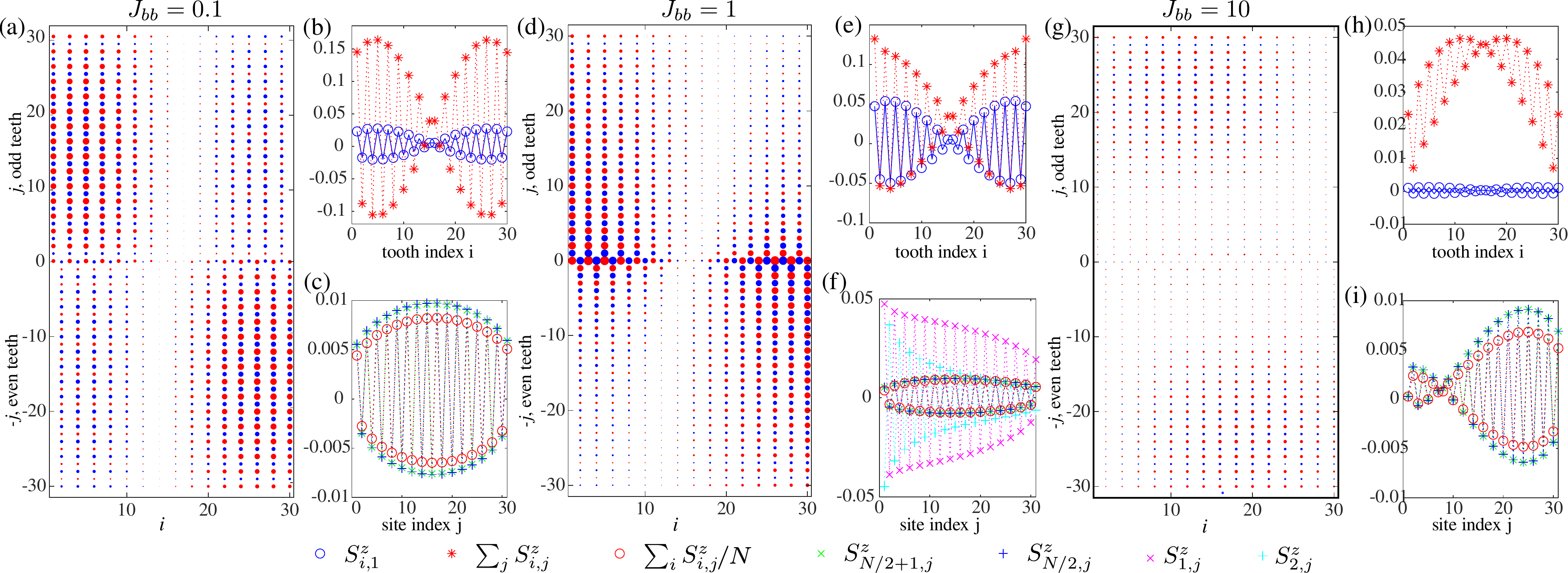}
\caption{(Color online) Same as Fig.\ref{fig:locmag_30_30} but for odd $L=31$
and even $N=30$. In panel (f) in addition to the average magnetization and
central teeth profile, we also show the results on the first and second
teeth to highlight that the profiles change along the backbone. In panel
(g), with respect to panels (a) and (d), the size of circles were enlarged
by a factor of 2.} \label{fig:locmag_31_30}
\end{figure}

\begin{figure}[t!]
\centering
\includegraphics[width=0.75\textwidth]{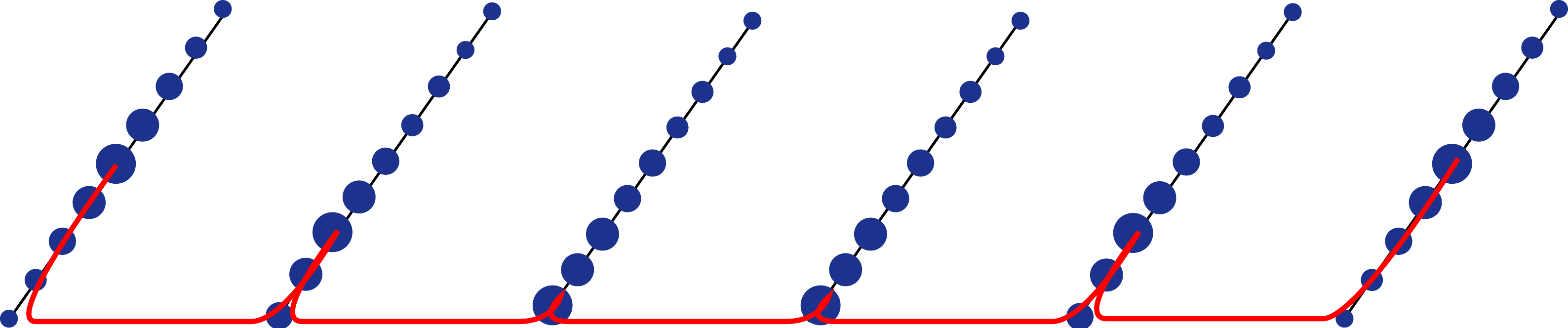}
\caption{(Color online) Schematic representation of a state in a comb with odd
number of sites per tooth and $J_{bb}\approx 1$. Spin-1/2 degrees of
freedom are delocalized along the teeth in a non-uniform way, so the emergent spin-1/2 degrees of freedom are not equally spaced. This is effectively equivalent to the spin-1/2 chain with non-uniform coupling constant $J_i\neq$const. }
\label{fig:delocalized2}
\end{figure}

Let us now look at the case of $L$-odd and $N$-even. In the weak-backbone
regime, each tooth is in a state with total spin-1/2, so along the backbone we
observe a critical spin-1/2 chain with a pronounced butterfly profile of local
magnetization shown in Fig.\ref{fig:locmag_31_30}(b). Importantly, the location
of the maximum (or minimum) of magnetization on different teeth is almost the
same, although the values at the maximum are very different (see
Fig.\ref{fig:locmag_31_30}(a),(c)), which qualitatively agrees with our picture
sketched in Fig.\ref{fig:delocalized} with the uniform lattice spacing of an
effective spin-1/2 chain. By contrast, at $J_{bb}=1$ the location of maxima on
different teeth are different (see Fig.\ref{fig:locmag_31_30}(d),(f)). So the
distance between the effective spin-1/2 degrees of freedom is not uniform, as
sketched in Fig.\ref{fig:delocalized2}. 
As a result, the finite-size
scaling of the energy gap is not linear neither with
$1/L$(Fig.\ref{fig:gap_1}(c)), nor $1/N$(Fig.\ref{fig:gap_1}(d)), nor with the
product of the two. Moreover, the larger $N$ and $L$ we take the more freedom
(or disorder) we add to the system, so less conformal the scalings are.
This agrees with the fact that both values $d_N$ and $d_L$ move away from the
CFT-invariant dimension $d_{N,L}=1$.
In a critical 1D system, spin-spin correlations decay with the distance between the spins as a power-law; therefore it is natural to expect that the non-uniform distribution of the spin-1/2 degrees of freedom leads to a non-uniform coupling constant in an effective spin-1/2 chain.

\begin{figure}[t!]
\centering
\includegraphics[width=1\textwidth]{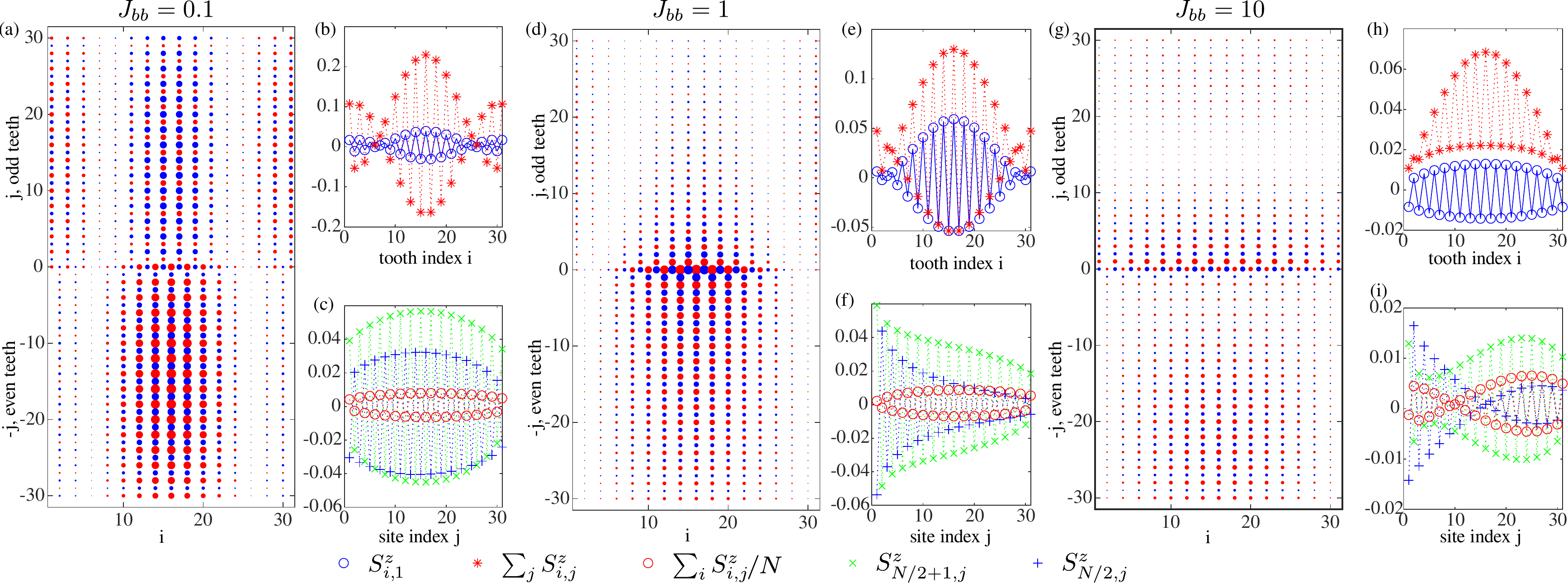}
\caption{(Color online) Same as Fig.\ref{fig:locmag_30_30} but for odd
$L=N=31$. Here we plot the difference in magnetization between the lowest
energy state in the sector of $S^z_\mathrm{tot}=3/2$ and the ground-state
in the sector of $S^z_\mathrm{tot}=1/2$. In panel (g), with respect to
panels (a) and (d), the size of circles were enlarged by a factor of 2.}
\label{fig:locmag_31_31}
\end{figure}

The same type of argument can be applied to a comb with both $N$ and $L$ odd.
At $J_{bb}=0.1$ effective spin-1/2 degrees of freedom are equally distant
from each other as can be deduced from Fig.\ref{fig:locmag_31_31}(a). However,
the equidistance is destroyed at $J_{bb}=1$ (see Fig.\ref{fig:locmag_31_31}
(d),(f)) and the finite-size scaling of the gap is non-linear
(Fig.\ref{fig:gap_1}(f),(g)).

\section{Large-backbone limit}

\begin{figure}[t!]
\centering
\includegraphics[width=0.5\textwidth]{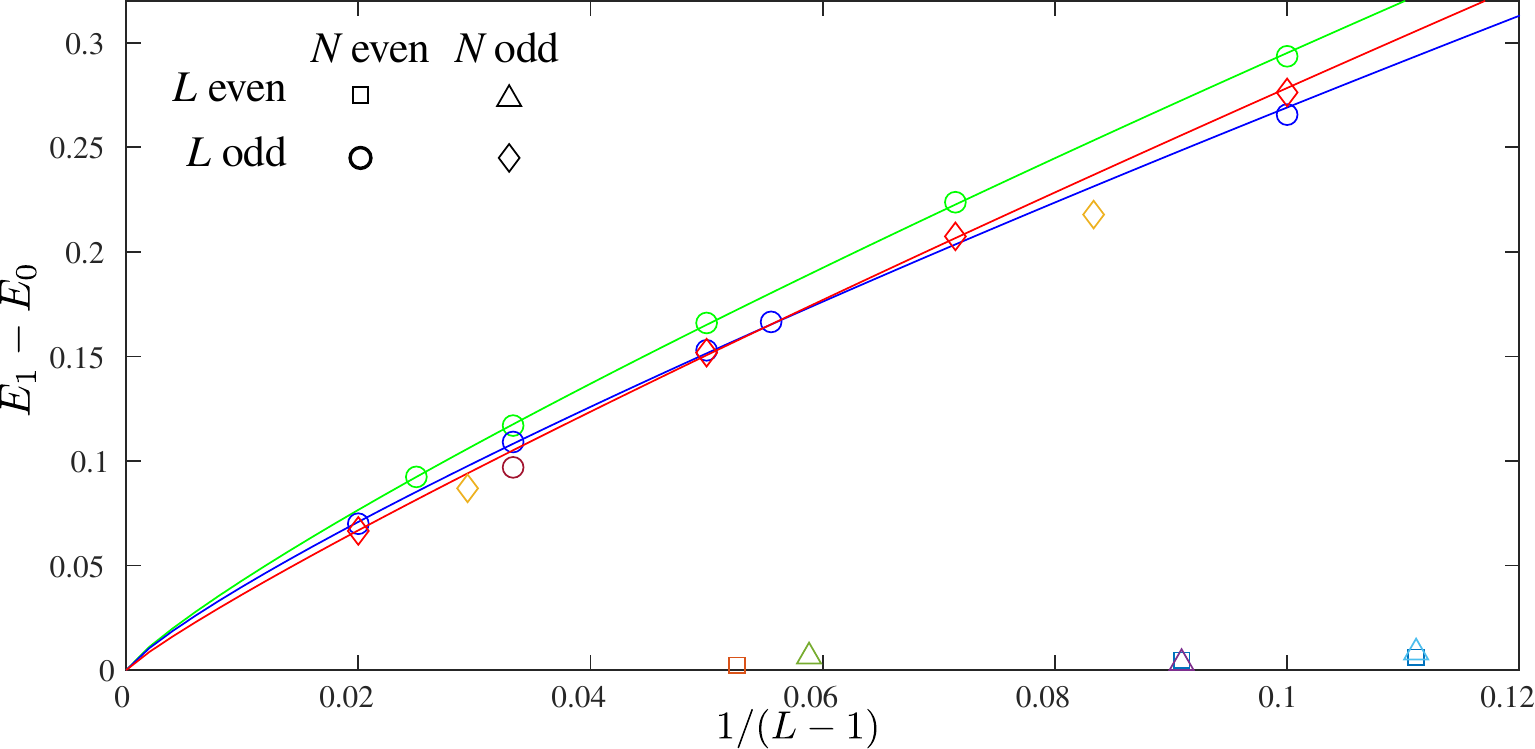}
\caption{(Color online) Scaling of the lowest magnetic gap in a strong-backbone
regime. Very low excitation energy (of the order of $10^{-3}$) is detected
for $L$ even (squares and triangles). For odd $L$ the finite-size gap is
much larger, but vanishes in the thermodynamic limit. Lines are the result
of the fit of equal-$N$ data to $\propto (L-1)^{-d_L}$. The resulting values of the
scaling dimension are within the range $0.83<d_N<0.89$.}
\label{fig:gapstrong}
\end{figure}

Finally, let us consider the strong backbone limit $J_{bb}\gg 1$. 
According to Fig.\ref{fig:graph}(e)-(f), when the backbone interaction is
sufficiently large, it is almost decoupled from the remaining teeth, each of
which becomes one site shorter. As we know from the discussion an even-odd
effect plays a crucial role. So let us start with  $L$ even.  The shortening of
the teeth is  confirmed by the anti-symmetric  profile of the Friedel oscillations
particularly for chains with an odd number of sites, but also observed in a comb with even
$N=L=30$ at $J_{bb}=10$ (see Fig.\ref{fig:friedel}(a)). As the result, each
tooth is in a state with total spin-1/2, weakly coupled to the strong
backbone. The entire picture resembles the story of a two-leg ladder, which
remains in the rung-dimer phase for any value of leg coupling constant. However
the absence of the coupling along the "second leg" of the ladder implies that
dangling spins can be polarized at very low energy. On various clusters from
size $8\times 8$ up to $20\times 10$, for which we could reach the convergence,
the gap is of the order of $10^{-3}$ (see Fig.\ref{fig:gapstrong}). 

The structure of the lowest excitations can be deduced from
Fig.\ref{fig:locmag_30_30}(g)-(i) and Fig.\ref{fig:locmag_30_31}(g)-(i). The
triplet state corresponds to the two excited teeth far apart from each other.
Interestingly enough, for both - even and odd $N$ - the maximum of
magnetization is allocated at the second and the $N-1$'s teeth.  For some
reason these excitations avoid the first and the last teeth when $J_{bb}\gg 1$.

In case of odd $L$ the system corresponds to the critical backbone chain which
is almost decoupled from the teeth with even $(L-1)$ number of particles. By
analogy with the previous even-$L$ case, this can be viewed as a critical chain
decorated by spin-0 objects. Because of the difference in couplings, it is
energetically favorable to accommodate a triplet excitation on a tooth than to
excite the "heavy" backbone. However, since the backbone itself is critical the
excited tooth is delocalized as shown in Fig.\ref{fig:locmag_31_30}(g)-(i). 
The scaling of the excitation energy as in the case of an isolated chain is affected by the logarithmic corrections of the form $\propto -\frac{\pi v}{(L-1)\log (L-1)}$, which reduce apparent scaling dimension obtained from the numerical fit to $E_1-E_0=\pi v/(L-1)^{d_\mathrm{app}}$, from its CFT value $d=1$. 

The
picture is a bit more complicated when both $L$ and $N$ are odd, since the
backbone itself is in the spin-1/2 state, which couples to a triplet excitation
on a tooth and gives complicated structure shown in
Fig.\ref{fig:locmag_31_31}(g)-(i).

\section{Discussion}

To summarize, we have studied numerically the Heisenberg spin-1/2 model on a
comb lattice. We have found very different properties of the low-energy states
depending on whether the number of sites per tooth is even or odd. The observed
even-odd effect is similar to that of spin-1/2 ladders with even and odd
number of legs\cite{Dagotto}.

\begin{figure}[t!]
\centering
\includegraphics[width=1\textwidth]{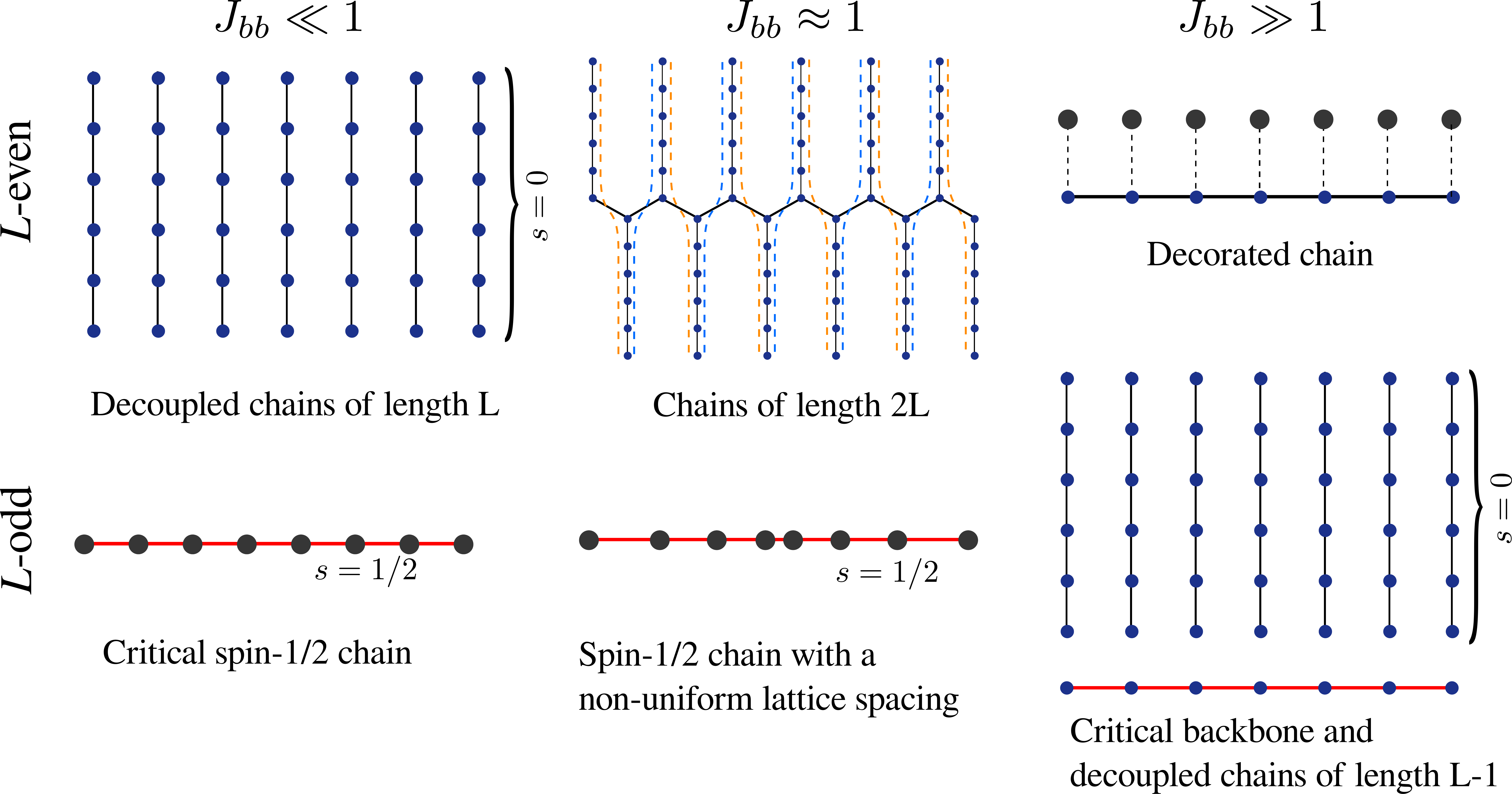}
\caption{(Color online) Summary of the main critical regimes as a function
backbone interaction and total number of spins per tooth. Non-uniform lattice spacing for $L$-odd is equivalent to the non-uniform coupling constant $J_i$ that varies along the chain.}
\label{fig:summary}
\end{figure}

In each case we detect three main regimes while tuning the backbone
interaction. They are summarized in Fig.\ref{fig:summary}.  When $L$ is even, by
changing the backbone interaction one can interpolate between nearly-decoupled
chains of length $L$ to an extended chains of length $2L$ when backbone and
teeth couplings are competing. Finally, in the strong-backbone limit the system
corresponds to a two-leg ladder with zero-coupling along one leg or to the
decorated spin-1/2 chain. The decorating spins are spin-1/2 degrees that
corresponds to the ground-state of teeth that contains odd $L-1$ sites each.

For odd $L$ the system corresponds to the critical  chain with very unusual
$1/(NL)$ finite-size scaling of the spectrum due to delocalization of the
spin-1/2s along the teeth. Tuning the backbone interaction effectively changes
the lattice spacing in an effective spin chain. When the backbone interaction
is comparable to the coupling along the teeth the effective spins are placed
along the chain in a non-uniform way so the conformal invariance of the system
in its usual sense is destroyed. However this opens an important question for
conformal field theory: how the two CFTs in 1+1D interact with each other; and
how can one  describe a resulting effect of competing criticalities? In the
present case the competition was induced by the chosen geometry of the lattice
and naturally leads to the competition between the two CFTs. However, a similar
scenario can be expected also on 2D lattices if rotation symmetry is
spontaneously broken, e.g. in case of helical or stripe states close to or at
the critical point.  The answer to these questions lies far beyond the scope of
this manuscript, however we hope that the present work will stimulate further
theoretical studies in this direction.

\section{Acknowledgements}
We  acknowledge  insightful  discussions  with  Ian  Affleck. This work has been supported by the Swiss National Science Foundation and the U.S. National Science Foundation
through DMR-1812558, and the Simons Foundation through the Many-Electron Collaboration. The  calculations  have  been  performed using the  computing facilities of the University of Amsterdam.

\bibliography{bibliography}

\nolinenumbers

\end{document}